\documentstyle [12pt] {article}

\newcommand{\h}[1]{\mathop{\lambda}\limits_{#1}\ \!\!\!}

\newcommand{\edf}{\ {\mathop{=}\limits^{\rm def}}\ }

\newcommand{\al}{\alpha}
\newcommand{\be}{\beta}
\newcommand{\m}{\mu}
\newcommand{\n}{\nu}
\newcommand{\s}{\sigma}

\begin{document}
\title{\bf Absolute Parallelism Geometry: Developments, Applications and Problems}

\author{{\bf M. I. Wanas}\\
\normalsize Astronomy Department, Faculty of Science, Cairo
university, Egypt.\\ e. mail wanas@frcu.eun.eg} \maketitle
\begin{abstract}

Absolute parallelism geometry is frequently used for physical
applications. It has two main defects, from the point of view of
applications. The first is the identical vanishing of its
curvature tensor. The second is that its autoparallel paths do not
represent physical trajectories. The present work shows how these
defects were treated in the course of development of the geometry.
The new version of this geometry contains simultaneous
non-vanishing torsion and curvatures. Also, the new paths
discovered in this geometry do represent physical trajectories.
Advantages and disadvantages of this geometry are given for each
stage of its development. Physical applications are just mentioned
without giving any details.
\end{abstract}
\section{Introduction}

  {\it "To understand {\bf nature}, one has to start with {\bf geometry} and end with
    {\bf physics}"} \\
This statement summarizes the {\bf geometrization} philosophy,
 introduced by Albert Einstein at the beginning of the
20th century. Einstein succeeded in applying this philosophy to
construct his theory of general relativity (GR). He started with
Riemannian geometry and ended with a successful theory for
gravity.

After the success of this theory, Einstein tried to construct a
theory unifying gravity and electromagnetism, using the same
philosophy. He realized the fact that Riemannian Geometry is not
the appropriate  candidate for his aim. This geometry is
relatively limited and it is just sufficient to describe gravity
alone. It has only a unique affine connection, a unique curvature
tensor and two path equations (the geodesic and the null
geodesic). The building blocks of this geometry are ten components
of the metric tensor (for n=4) which are just sufficient to
describe gravity as stated above. So, he started to look for
another geometry, wider than the Riemannian one. Einstein started
his first attempt in this context in 1928 by using Absolute
Parallelism (AP) geometry (some authors prefer to call it distant
parallelism or teleparallelism or fernparallelismus). Due to the
long history of development of this geometry, about seventy five
years, a full review of its applications and developments cannot
be given in such a small number of pages. In what follows, the
main lines of this subject are discussed in four stages. Readers
interested in the details are referred to the references listed at
the end of this paper.

\section{ The First Stage (1928-1951) }

 In this stage the geometric structure of the AP-space can be
 summarized as follows (cf. [1],[2]):
 An absolute parallelism space is an n-dimensional manifold each
 point of which is labeled by n-independent variables $x^{\nu} (\nu =1,2,3,...n)$
 and at each point we define n-linearly independent contravariant vectors
$\h{i}^\mu (i=1,2,3,...,n$, denotes the vector number and $\mu
=1,2,3...n$ denotes the coordinate component) subject to the
condition, $$
 \h{i}^\mu_{.|~ \nu}=0,  \eqno{(2.1)}
$$
 where the stroke denotes absolute differentiation to be
 defined later. Equation (2.1)is the condition for the absolute
 parallelism.
 The covariant components of $\h{i}^\mu$ are defined such that:
 $$
\h{i}^\mu \h{i}_\nu = \delta^{\mu}_{\nu}, \eqno{(2.2)}
 $$
and
 $$
  \h{i}^{\nu} \h{j}_{\nu} = \delta_{ij}.  \eqno{(2.3)}.
$$ Using these vectors, the following second order symmetric
tensors are defined: $$ g^{\mu \nu} \edf \h{i}^\mu \h{i}^{\nu},
\eqno{(2.4)} $$

$$ g_{\mu \nu} \edf \h{i}_\mu \h{i}_{\nu}, \eqno{(2.5)} $$
consequently,
 $$ g^{\mu \alpha} g_{\nu \alpha} = \delta^{\mu}_{\ \nu }. \eqno{(2.6)} $$
These second order tensors can serve as metric tensors of
Riemannian space, associated with the AP-space, when needed. This
type of geometry admits, at least, four affine connections. The
first is non-symmetric connection given as a direct solution of
the AP-condition, i.e. $$ \Gamma^{\alpha}_{.~\mu \nu} =
\h{i}^{\alpha} \h{i}_{\mu,\nu}. \eqno{(2.7)} $$ The second is its
dual $ \hat{\Gamma}^{\alpha}_{.~\mu \nu}( = \Gamma^{\alpha}_{.~\nu
\mu}) $ since (2.7) is non-symmetric. The third one is the
symmetric part of (2.7),  $ \Gamma^{\alpha}_{.(\mu \nu)}$. The
fourth is Christoffel symbol defined using (2.4),(2.5) ( as a
consequence of a metricity condition of the associated Riemannian
space). The torsion tensor[3] is twice the skew symmetric part of
the affine connection (2.7), i.e. $$ \Lambda^{\alpha}_{.~\mu \nu}
\edf \Gamma^{\alpha}_{.~\mu \nu} - \Gamma^{\alpha}_{.~\nu \mu}.
\eqno{(2.8)} $$ Another third order tensor (contortion) is defined
by, $$ \gamma^{\alpha}_{.~\mu \nu} \edf \h{i}^{\alpha} \h{i}_{\mu
; \nu}, \eqno{(2.9)} $$ where the semicolon is used for covariant
differentiation using Christoffel symbol. The two tensors are
related by, $$\gamma^{\alpha}_{.\mu \nu}= \frac{1}{2}
(\Lambda^{\alpha}_{.\mu \nu } - \Lambda^{~ \alpha}_{\nu .\mu} -
\Lambda^{~\alpha}_{\mu .\nu}). \eqno{(2.10)}
 $$

 A basic vector could be
 obtained by contraction of the above tensors,
$$ C_{\mu} \edf  \Lambda^{\alpha}_{.\mu \alpha }=
\gamma^{\alpha}_{. \mu \alpha}. \eqno{(2.11)}$$ One of the
advantages of AP- geometry is that for any tensor
$T^{\alpha}_{.~\beta \gamma}$ defined in the AP-space one can
construct a set of scalars $T_{ijk}$,

$$ T_{ijk} \edf \h{i}_{\alpha} \h{j}^{\beta} \h{k}^{\gamma}
T^{\alpha}_{.~ \beta \gamma}. \eqno{(2.12)}$$ If
$T^{\alpha}_{.~\beta \gamma}$ is the contortion (2.9) then the
corresponding scalars are known in the literature as Ricci
coefficients of rotation [4].

 The curvature tensor, in this stage, is defined by,
 $$
B^{\al}_{.\m \n \s} {\ }  \edf {\ } \Gamma^{\al}_{.\m  \s, \n} -
\Gamma^{\al}_{.\m \n, \s} + \Gamma^{\al}_{\epsilon \n} \Gamma^
{\epsilon}_{. \m \s} - \Gamma^{\al}_{. \epsilon \s}
\Gamma^{\epsilon}_{. \m \n} \equiv 0.\eqno{(2.13)}
 $$
This tensor vanishes identically because of (2.1). The
autoparallel path equation can be written in the form, $$
\frac{d^{2}x^{\mu}}{d \lambda^{2}}+ \Gamma^{\mu}_{\alpha \beta}
\frac{d x^{\alpha} }{d \lambda}\frac{d x^{\beta} }{d \lambda} =
0.\eqno{(2.14)} $$

Robertson in 1932 developed the theory of groups of motion in
AP-spaces [5]. He constructed three AP-structures, one with
spherical symmetry and the other two with homogeneity and isotropy
suitable for cosmological applications. These structures are
important for physical applications.

 The only application of this
geometry was Einstein's series of papers e.g.[3],[6], to construct
a field theory, in an attempt to unify gravity with
electromagnetism. Unfortunately, this attempt failed for the
following reasons. The first reason is that the solution of the
unified field equations in the case of spherical symmetry [6] did
not reduce to the Schwarzschild solution when electromagnetism was
switched off. The second reason is the non-physical consequence of
axially symmetric solution of theory [7]. The third reason was
that autoparallel paths do not represent physical trajectories of
any test particle (cf.[8]). It is worth of mention that
Levi-Civita had tried to simplify Einstein's unified field
equations, but he really wrote a different theory [4].

 It may be
of interest to give a brief comparison between the AP-geometry and
the Riemannian one in this stage. Table 1 summarizes the main
features of each geometry. The objects displayed in this table are
those important for applications. It is clear from this table that
the AP-geometry is more wider, from the point of view of
applications, than the Riemannian one. For example, for $n=4$, the
number of independent components of the building blocks of
Riemannian geometry is ten, which is just sufficient to describe
gravity. On the other hand, the corresponding number in the
AP-geometry is sixteen. So, the extra degrees of freedom of the
AP-geometry  could be used to represent other physical fields together with
gravity.

 The above lines give a brief review of AP-geometry in its first stage.

\begin{center}
 Table 1: Comparison Between The Riemannian Geometry and
 AP-Geometry \\
\vspace{0.5cm}
\begin{tabular}{|c|c|c|} \hline
& & \\ Object  & Riemannian geometry & AP-geometry
\\ & &
\\ \hline & & \\ Building Blocks     & $g_{\mu\nu} $
& $\h{i}^{\mu}, $
\\ & & \\ \hline & & \\ Affine Connection & $ \{^{\m}_{\al\be}\}$   &
$ \{^{\m}_{\al\be}\},
\Gamma^{\alpha}_{\mu\nu},\hat\Gamma^{\alpha}_{\mu\nu}$,
$\Gamma^{\alpha}_{(\mu\nu)}$
\\ & &
\\ \hline & &
\\ Second Order Symmetric Tensors
 & $g_{\mu\nu},R_{\mu\nu}$ &
$g_{\mu\nu},R_{\mu\nu} $\\ & & \\ \hline & &
\\ Second Order Skew Tensors &
---- & $\xi_{\mu\nu},\zeta_{\mu\nu}$ \\ & & \\ \hline & & \\
 Third Order Tensor & ---- & $\gamma^{\alpha}_{.\mu\nu},\Lambda^{\alpha}_{.\mu\nu} $ \\
  & & \\ \hline & &\\
Vectors & ---- & $C_{\mu}$
\\ & & \\ \hline & & \\ Scalars & R & Many
\\ & & \\ \hline
 & & \\
Curvature& $R^{\alpha}_{.\beta\gamma\delta} \neq 0$ &
$B^{\alpha}_{.\beta\gamma\delta} \equiv  0$ \\ & & \\ \hline
\end{tabular}
\end{center}
    In this stage the AP-geometry has two
main problems concerning applications: The first is the identical
vanishing of its curvature tensor and the second is that its path
equations do not represent physical trajectories. After the
failure of Einstein's attempt, the AP-geometry was neglected for
about twenty years except one or two papers by A.G. Walker in
1940's, e.g. [9], obtaining results similar to those of Robertson.

\section{The Second Stage (1952-1974)}

In this stage the development of AP-geometry can be summarized in
the following. Mikhail in 1952 [10] revisited this geometry and
constructed the following second order tensors (Table 2) which are
important for applications, with the algebraic identity, $$
\eta_{\mu \nu} + \varepsilon_{\mu \nu} - \xi_{\mu \nu} \equiv 0.
\eqno{(3.1)}  $$
\eject
\begin{center}
 Table 2: Second Order World Tensors [10]      \\
\vspace{0.5cm}
\begin{tabular}{|c|c|} \hline
 & \\
Skew-Symmetric Tensors                &  Symmetric Tensors   \\
 & \\ \hline
 & \\
${\xi}_{\mu \nu} \edf \gamma^{~ ~ \alpha}_{\mu \nu .
|{\stackrel{\alpha}{+}}} $ &
\\

${\zeta}_{\mu\nu} \edf C_{\alpha}~{\gamma^{~~ \alpha}_{\mu \nu .}
} $ &
\\
 & \\ \hline
 & \\
${\eta}_{\mu \nu} \edf C_{\alpha}~{\Lambda^{\alpha}_{.\mu \nu} } $
& ${\phi}_{\mu \nu} \edf C_{\alpha}~\Delta^{\alpha}_{.\mu \nu} $
\\

${\chi}_{\mu \nu} \edf \Lambda^{\alpha}_{. \mu
\nu|{\stackrel{\alpha}{+}} }$ & ${\psi}_{\mu \nu} \edf
\Delta^{\alpha}_{. \mu \nu|{\stackrel{\alpha}{+}}} $
\\

${\varepsilon}_{\mu \nu} \edf C_{\mu | {\stackrel{\nu}{+}}} -
C_{\nu | {\stackrel{\mu}{+}}}$ & ${\theta}_{\mu \nu} \edf C_{\mu |
{\stackrel{\nu}{+}}} + C_{\nu | {\stackrel{\mu}{+}}}  $
\\

${\kappa}_{\mu \nu} \edf \gamma^{\alpha}_{. \mu
\epsilon}\gamma^{\epsilon}_{. \alpha \nu} - \gamma^{\alpha}_{. \nu
\epsilon}\gamma^{\epsilon}_{. \alpha \mu}$   & ${\varpi}_{\mu \nu}
\edf  \gamma^{\alpha}_{. \mu \epsilon}\gamma^{\epsilon}_{. \alpha
\nu} + \gamma^{\alpha}_{. \nu \epsilon}\gamma^{\epsilon}_{. \alpha
\mu}
$
\\
 & \\ \hline
 & \\
                  &  ${\omega}_{\mu \nu} \edf \gamma^{\epsilon}_{. \mu \alpha}\gamma^{\alpha}_{. \nu \epsilon}$   \\

                                      &  ${\sigma}_{\mu \nu} \edf \gamma^{\epsilon}_{. \alpha \mu} \gamma^{\alpha}_{. \epsilon \nu}$   \\

                                      &  ${\alpha}_{\mu \nu} \edf C_{\mu}C_{\nu}$   \\

                                      &  $R_{\mu \nu} \edf \frac{1}{2}(\psi_{\mu \nu} - \phi_{\mu \nu} - \theta_{\mu \nu}) + \omega_{\mu \nu}$          \\
 & \\ \hline
\end{tabular}
\end{center}
Where $\Delta^{\alpha}_{. \mu \nu}$ is twice the symmetric part of
the contorsion (2.9),$R_{\mu \nu}$ is Ricci tensor and $T_{\mu \nu
|\stackrel{\sigma}{+}} \equiv
T_{\stackrel{\mu}{+}{\stackrel{\nu}{+}} | \sigma}$.

 Hayashi and
Bragman [11] derived the irreducible decomposition of torsion
tensor which can be written as, $$ \Lambda_{\alpha \mu \nu} =
\frac{2}{3} (t_{\alpha \mu \nu} - t_{\alpha \nu \mu}) +
\frac{1}{3}(g_{\alpha \mu }C_{\nu} -g_{\alpha \nu}C_{\mu}) +
\epsilon_{\alpha \mu \nu \sigma}a^{\sigma},\eqno{(3.2)} $$ where
$$ t_{\alpha \mu \nu} \edf \frac{1}{2} (\Lambda_{\alpha \mu \nu} +
\Lambda_{\mu \alpha  \nu}) + \frac{1}{6}(g_{\nu \alpha}C_{\mu} +
g_{\mu \nu }C_{\alpha}) - \frac{1}{3}g_{\alpha
\mu}C_{\nu},\eqno{(3.3)}$$ $$
 a_{\mu} \edf \frac{1}{6}~{\epsilon}_{\mu \alpha \beta
 \gamma}~\Lambda^{\alpha \beta \gamma},\eqno{(3.4)}$$
 and  $ {\epsilon}_{\mu \alpha \beta\gamma}$ is the Levi-Civita
 tensor.

  Although the development of the AP-geometry in this stage was
not as big as in its first stage, its applications were carried
out in many diverse problems. For example, McCrea and Mikhail
[10],[12], have used this geometry to modify GR in order to
account for continuous creation of matter in the Universe. Bergman
and Thomson [13] have used it to treat spin and angular momentum
in GR. Bilby et al. [14] have used the geometry in studying
dislocations in crystals. Utiyama [15], Kibble [16] and Sciama
[17] started some attempts for gauging gravity using AP-geometry.
M\o ller [18], [19] has used the geometry to solve the problem of
localization of energy in GR, a problem that is impossible to be
solved in the framework of Riemannian geometry. Mikhail [20] has
constructed a pure geometric unified field theory using the
AP-geometry. Hehl [21] continued the attempts of Utiyama, Kibble
and Sciama to construct a gauge field theory for gravity.

\section{The Third Stage (1975-1994)}

Many authors believe that, because of (2.13), the AP-space is a
flat one. In this stage it is shown that AP-spaces are, in
general, curved. The problem of curvature in AP-spaces is a
problem of definition. In any affinely connected space there are,
at least, two methods for defining the curvature tensor. The first
method is by replacing Christoffel symbol, in the definition of
curvature tensor of Riemannian geometry, by the connection
defined in the space. The second method is to define curvature as
a measure of non-commutation of covariant (absolute in the present
case) differentiation using the connection of the space. It
is known that, the two methods give identical results in case of
Riemannian space. But the situation is different for spaces with
non-symmetric connections. The two methods are not identical.

 The application of the second method
for non-symmetric geometries implies a problem. That is, we
usually use an arbitrary vector in order to study the
non-commutation of covariant differentiation, and the resulting
expression is not free from this vector. Fortunately, this problem
is solved for the AP-spaces [22]. We can replace the arbitrary
vector by the vectors defining the structure of AP-spaces. In this
case we can define the following curvature tensors:

$$ \h{i}^{\stackrel{\m}{+}} _{\  | {\n \s}}  -
\h{i}^{\stackrel{\m}{+}}_ {\  | {\s \n}} \edf   \h{i}^\al B^\m_{.
\al \n \s},\eqno{(4.1)} $$ $$ \h{i}^{\stackrel{\m}{-}}_ {\ | {\n
\s}} - \h{i}^{\stackrel{\m}{-}}_{\  | {\s \n}}  \edf  \h{i}^\al
L^\m_{. \al \n \s},\eqno{(4.2)} $$ $$ \h{i}^{\stackrel{\m}{}}_ {\
| {\n \s}}  - \h{i}^{\stackrel{\m}{}} _{\ | {\s \n}} \edf
\h{i}^\al N^\m_{. \al \n \s},\eqno{(4.3)} $$
 $$
\h{i}^{\mu}_{\  ; \n \s}  - \h{i}^{\mu}_{\  ; \s \n}  \edf
\h{i}^\al R^\m_{. \al \n \s} ,\eqno{(4.4)}  $$ here we use the
stroke , a (+) sign  and (-) sign to characterize absolute
differentiation using the connection (2.7) and its dual,
respectively. We use the stroke without signs to characterize
absolute differentiation using the symmetric part of (2.7), while
the semicolon is used to characterize covariant differentiation
using the Christoffel symbols. The curvature tensors defined by
(4.1), (4.2), (4.3) and (4.4) are in general non-vanishing except
the first one, which vanishes (because of the AP-condition). From
the application point of view, one of the two problems  of the
AP-geometry, the curvature problem, is partially solved. There is
a net of relations between the different contractions of these
tensors [22]. Also there is another net of relations between the
absolute derivative of different geometric objects.

A Lagrangian, built using these curvature tensors, has been used
to construct a field theory unifying gravity and electromagnetism
[22], [23]. When this theory was linearized both Newton's theory
of gravitation and Maxwell's theory of electromagnetism were
obtained.Also an interpretation of Lorentz condition , used
usually in solving Maxwell's equations, is given [24].

Furthermore, a covariant scheme for classifying AP-spaces, known
as{\it "type analysis"}, is suggested [24]. To clarify the
physical meaning of the type analysis we give a trivial example
from Riemannian geometry. It is well known that gravity cannot be
represented, properly, in flat spaces. So, if we have a Riemannian space with
a certain metric and we want to construct a model for gravity
using this metric, it is better first to calculate the corresponding curvature
tensor for this metric. The result is either a vanishing curvature
tensor,or a non-vanishing one. Then, we can classify Riemannian
spaces, from the application point of view, to two classes $G0$
and $GI$, say. The first $(G0)$ is the class with vanishing
curvature, which is not appropriate for application concerning gravity. The second
$(GI)$ is the class with non-vanishing curvature and is the good
candidate for application. The same idea is applied in the case of
AP-spaces using other tensor together with the curvature one.
Table 3 gives a summary of this classification. It is to be
considered that this classification scheme "{\it{the type
analysis}}" is a covariant one, i.e. independent of coordinate
system used. Using the type analysis, one can know, before solving
the field equations, the type and strength of the fields that an
AP-space is capable of representing.
\begin{center}
 Table 3: Type Analysis       \\
\vspace{0.5cm}
\begin{tabular}{|c|c|c|} \hline
& & \\ Indicator  & Field Represented  & Type
\\ & &
\\ \hline & & \\ $F_{\mu\nu}= 0$     & No electromagnetic field. & $F0$
\\ & & \\ \hline & & \\ $F_{\mu\nu} \neq 0 $, $\zeta_{\mu\nu} =0 $ &  Weak electromagnetic field.   &
$FI$ \\ & & \\ \hline & & \\ $F_{\mu\nu} \neq 0 $, $\zeta_{\mu\nu}
\neq 0$  & Strong electromagnetic field. & $FII $\\ & & \\ \hline
& &
\\ $R^{\alpha}_{.\beta\gamma\delta} = 0$  & No gravitational
field. & $G0$ \\ & & \\ \hline & & \\
$R^{\alpha}_{.\beta\gamma\delta} \neq 0$, $T_{\mu\nu} = 0$,
$\Lambda = 0$ & Weak gravitational field in free space. & $GI$ \\
& & \\ \hline & &\\ $R^{\alpha}_{.\beta\gamma\delta} \neq 0$,
$T_{\mu\nu} \neq 0$, $\Lambda = 0$ & Gravitational field within a
material distribution. & $GII$ \\ & & \\ \hline & & \\
$R^{\alpha}_{.\beta\gamma\delta} \neq 0$, $T_{\mu\nu} \neq 0$,
$\Lambda \neq 0$ & Strong gravitational field within a material
distribution. & $GIII$
\\ & & \\ \hline
\end{tabular}
\end{center}
As stated before, the AP-geometry has extra (six) degrees of
freedom more than the Riemannian geometry. If these extra degrees
of freedom are attributed to electromagnetism $(F)$ [23], [24],
and the other ten degrees of freedom are used to represent gravity
$(G)$, then possible combinations between $G$ and $F$ classes
could be listed in the following two groups:
\\ The first group: $F0G0, F0GI, F0GII, FIGII$. \\ The second
group: $ F0GIII, FIGIII, FIIGII, FIIGIII.$ \\ It is shown that
applications using models from the first group give good agreement
with classical field theories of gravitation and electromagnetism
[25], [26]. Deviation from classical field theories appear when
using models from the second group (e.g. [27]). Tensors used for
classifications are combinations of tensors given in Table 2. viz,
$$F_{\mu\nu} \edf Z_{\mu\nu} - \xi_{\mu\nu}, $$ $$Z_{\mu\nu} \edf
\eta_{\mu \nu} + \zeta_{\mu\nu },$$ $$ T_{\mu\nu} \edf
g_{\mu\nu}\Lambda + \varpi_{\mu\nu} - \sigma_{\mu\nu}, $$
$$\Lambda \edf \frac{1}{2}(\sigma - \varpi) $$

 Hehl et al. [28] have used the AP-geometry in constructing a
gauge theory for gravity considering the Poincar{\'e} group. M\o
ller in 1978 [29] tried to overcome the singularity problem of GR
by using the AP-geometry. Hayashi and Shirafuji in 1979
constructed a microscopic gauge field theory for gravity
considering the translation group [30].

 The curvature tensors
defined above can be written explicitly in terms of torsion or
contortion via (2.10), i.e.  $$ B^{\alpha}_{. \mu \nu \sigma} =
R^{\alpha}_{. \mu \nu \sigma} + Q^{\alpha}_{. \mu \nu \sigma}
\equiv 0 ,\eqno{(4.5)} $$ $$
 L^{\alpha}_{. \mu \nu \sigma} = \Lambda^{\stackrel{\alpha}{+}}_{. {\stackrel{\m}{+}} {\stackrel{\nu}{-}}
 | \sigma} - \Lambda^{\stackrel{\alpha}{+}}_{. {\stackrel{\m}{+}} {\stackrel{\sigma}{-}}
 | \nu} + \Lambda^{\beta}_{. \mu \nu} \Lambda^{\alpha}_{. \sigma
 \beta} - \Lambda^{\beta}_{. \mu \sigma} \Lambda^{\alpha}_{. \nu
 \beta},
 \eqno{(4.6)}$$
 $$
N^{\alpha}_{. \mu \nu \sigma} = \Lambda^{\alpha}_{. \mu \nu |
\sigma } - \Lambda^{\alpha}_{. \mu \sigma | \nu } +
\Lambda^{\beta}_{. \mu \nu}\Lambda^{\alpha}_{.  \beta \sigma} -
\Lambda^{\beta}_{. \mu \sigma}\Lambda^{\alpha}_{.  \nu \beta},
\eqno{(4.7)}
 $$
$$ Q^{\alpha}_{. \mu \nu \sigma} =
\gamma^{\stackrel{\alpha}{+}}_{. {\stackrel{\m}{+}}
{\stackrel{\nu}{+}}
 | \sigma} - \gamma^{\stackrel{\alpha}{+}}_{. {\stackrel{\m}{+}} {\stackrel{\sigma}{-}}
 | \nu} + \gamma^{\beta}_{. \mu \sigma} \gamma^{\alpha}_{. \beta
 \nu} - \gamma^{\beta}_{. \mu \nu} \gamma^{\alpha}_{.  \beta \sigma},
 \eqno{(4.8)}
$$ It is clear that the vanishing of the torsion will lead to the
vanishing of (4.6), (4.7). Also this will lead to vanishing of
(4.8) via (2.10) and consequently the vanishing of $ R^{\alpha}_{.
\mu \nu \sigma}$ via (4.5). This is another defect of the geometry
which is connected to the viability of field theories written in
AP-spaces [31], [32]. This will be clarified in the next section.

\section{ The Fourth Stage (1995-2000)}

In this stage, new path equations were discovered in the
AP-geometry [33]. These equations can be written in the form: $$
{\frac{dU^\m}{dS^-}} + \{^{\m}_{\al\be}\} U^\al U^\be = 0,
\eqno{(5.1)}$$
 $$ {\frac{dW^\m}{dS^0}} + \{^{\m}_{\al\be}\} W^\al W^\be = -
{\frac{1}{2}} \Lambda^{~ ~ ~ ~ \m}_{(\al \be) .}~~ W^\al W^\be,
\eqno{(5.2)} $$ $$ {\frac{dV^\m}{dS^+}} + \{^{\m}_{\al\be}\} V^\al
V^\be = - \Lambda^{~ ~ ~ ~ \m}_{(\al \be) .} ~~V^\al V^\be.
\eqno{(5.3)} $$ This set of equations possesses some interesting
features: \\ 1. It gives the effect of the torsion on the curves
(paths)of the geometry. \\ 2. This set is irreducible i.e. no one
of these equations can be reduced to the other unless the torsion
vanishes. This will lead to flat space as mentioned at the end of
the last section.\\ 3. The coefficients of the torsion term jump
by a step of one-half from one equation to the next.\\ The last
feature is tempting to believe that {\bf "paths in this geometry
are naturally quantized"}.

As stated in section 2 the symmetric part of the connection (2.7)
is not Christoffel symbol. In some applications it is preferable
to have a non-symmetric connection whose symmetric part is the
Christoffel symbol. Such connection can be built by adding the
torsion to Christoffel symbols [34], $$ \Omega^{\alpha}_{. \mu\nu}
 \edf \{^{\alpha}_{\mu\nu}\} + \Lambda^{\alpha}_{. \mu \nu}.
\eqno{(5.4)} $$ This will add two affine connections to the
geometry of AP-spaces, (5.4) and its dual $ \hat
\Omega^{\alpha}_{.\mu \nu} \edf \Omega^{\alpha}_{.\nu \mu}$. Using
these connections we can define the following curvature tensors:
$$ \h{i}^{\stackrel{\m}{+}} _{\  || {\n \s}}  -
\h{i}^{\stackrel{\m}{+}}_ {\  || {\s \n}}   = \h{i}^{\al} M^\m_{.
\al \n \s},\eqno{(5.5)} $$ $$ \h{i}^{\stackrel{\m}{-}} _{\  || {\n
\s}}  - \h{i}^{\stackrel{\m}{-}}_ {\  || {\s \n}}   = \h{i}^\al
K^\m_{. \al \n \s}. \eqno{(5.6)} $$ We use the double stroke and a
(+) sign to characterize this type of absolute differentiation
using (5.4), and a (-) sign for its dual.

The situation now is that we have partially solved  the curvature
problem mentioned at the end of section 2. We have now defined six
curvature tensors, one of which vanishes identically while the
others do not. From the point of view of applications this
solution is partial since all these tensors could be written in
terms of the torsion tensor. Consequently, the vanishing of the
torsion will reduce the space to a flat one. For this property, it
is shown [35] that theories written in AP-spaces, in which the
torsion is connected with spin, will not be viable , since such
theories will not reduce to GR as the torsion vanishes. This
problem will be solved in the following lines.

The second problem of AP-geometry, mentioned in section 2 still
present. Although the new path equations, given above, have some
interesting features, these equations do not represent physical
trajectories of any particle. So, from the point of view of
applications, AP-geometry should be parameterized. Parameterizing
this geometry led to the following parameterized geometric objects
[36]. Combining linearly the above mentioned connections, we get
the following parameterized connection, $$ \nabla^{\mu}_{. \alpha
\beta } = (a+b) \{^{\mu}_{\alpha\beta}\} + b  \gamma^{\mu}_{.
\alpha \beta } \eqno{(5.7)} $$ where $a$ and $b$ are parameters.
As a consequence of metricity condition, using (5.7), we get
 $$ a + b = 1. $$
 The set of new paths (5.1), (5.2) and (5.3) can be generalized
 using (5.7). The result is
 the following parameterized path equation, [37]  $$ {\frac{dZ^\m}{d\tau}} + \{^{\m}_{\al\be}\} Z^\al
Z^\be = - b~~ \Lambda^{~ ~ ~ ~ \m}_{(\al \be) .} ~~Z^\al Z^\be,
\eqno{(5.8)} $$ where
 ${\tau}$ is a parameter varying along the path and ${Z^{\mu}}$ is the tangent to the path.
 All curvature tensors defined in this parameterized version of
 geometry, are non-vanishing. For example if we redefine the curvature (2.13) using the
 connection (5.7) we get [36] $$ \hat B^{\alpha}_{.\mu \nu \sigma} =  R^{\alpha}_{. \mu \nu \sigma} + b~~ \hat Q^{\alpha}_{. \mu \nu \sigma }.  \eqno{(5.9)}$$ This tensor is, in
 general non-vanishing although the corresponding one, in the
 old version of the geometry, vanishes identically.

\section{Concluding Remarks}
 The importance of the new version of AP-geometry, given in the last section, can be summarize in the following
 points: \\
 1. It is more general than the Riemannian geometry and the
 conventional
 AP-geometry . It has a general non-symmetric connection (5.7) giving rise to simultaneous non-vanishing curvature and
 torsion. This is in contrast to what mentioned by some authors, e.g. [38].\\
 2. If metricity condition is required  $(a + b = 1)$ then we can
 take either $a = 1 \Rightarrow b = 0$ and get Riemannian geometry without any need for a vanishing
 torsion, or we take $a = 0 \Rightarrow b = 1$ to   get the conventional
 AP-geometry. \\
 3. From the application point of view the parameter $b$ and the
 torsion term appearing in the path equation (5.8) have been
 connected to some physical interaction [37]. This equation has been used
to interpret [39] the discrepancy in the COW-experiment which
gives strong evidences for the existence of this interaction on
the laboratory scale. Another application [40] gives some
evidences for the existence of the interaction on the galactic
scale. A third application [41] studies the effect of this
interaction on the cosmological scale.
\\ 4. The parameterized version of AP-geometry is more suitable
for physical applications, especially for constructing theories
that require both torsion and curvature to describe different
interactions. For example attempts to geometrize strings [42],
theories accounting for Dirac fields [43] and  theories gauging
gravity [44], are among this class of theories.

The following table (Table 4) gives a summary of the important
developments, applications and problems of AP-geometry.

 \eject
\begin{center}
 Table 4: Stages of Developments, Applications and Problems of AP-geometry       \\
\vspace{0.5cm}
\begin{tabular}{|c|c|c|c|} \hline
& & &\\ Stage & Development & Applications  & Problems
\\ & & &
\\ \hline & & & \\
1928-1951 & Basic Structure.  & Gravity and electro- & Vanishing
curvature.
\\

           & Gps of motion in   & magnetism unification & Non physical
           paths.
\\ &AP-spaces. &[3]. & \\
& & &
\\ \hline & & & \\ 1952-1974 &  Second order   &
Unification [20].
 & Vanishing Curvature. \\ &  tensors and& Creation of
 matter [12]. & Non-physical paths. \\ &  identity. &  Isospin description [13]. & \\  &   &
  Gauging gravity [16]. & \\ & &  Material-energy  & \\ & & complex [19]. &
  \\
 & &Dislocations [14]. &
\\ & & & \\  \hline  & & &
\\ 1975-1994 &New absolute &Trials to quantize
& If $\Lambda^{\alpha}_{\mu\nu}=0\Rightarrow$ all
\\ &  derivatives. & gravity. & curvature tensors vanish. \\  & Non-vanishing & Unification [23]. &
Non-physical paths.\\
 & curvature tensors. & Gauging gravity [44]. &
\\  & Classification & Trials to solve  & \\ & of AP-Spaces. & the
singularity problem & \\ & & [29]. & \\ & & & \\ \hline & & & \\
1995-2000 & New paths & Delay of neutrinos & \\ & New affinity &
from SN1987A [40]. &
\\ & Parameterized & Interpretation of the & \\ & AP-geometry. &
discrepancy in the & \\ & & COW-experiment [39]. & \\ & &Quantum
paths [37]. &
\\ & & & \\ \hline
\end{tabular}
\end{center}

\section*{References}

{[1] Eisenhart, L.P. (1926) {\it "Riemannian Geometry"}, Princeton
Univ. Press.} \\ {[2] Eisenhart, L.P. (1927) {\it "Non-Riemannian
Geometry"}, American Math.Soc.} \\ {[3] Einstein, A. (1929) Sitz.
Preuss. Akad. Wiss., {\bf 1}, 1.} \\ {[4] Levi-Civita, T. (1929)
Sitz. Preuss. Akad. Wiss., {\bf 2}, 137.} \\ {[5] Robertson, H.P.
(1932) Ann. Math. Princeton (2), {\bf 33}, 496.} \\ {[6] Einstein,
A. and Mayer, W. (1930) Sitz. Preuss. Akad. Wiss., {\bf 1}, 110.}
\\ {[7] McVittie, G.C. (1930) Proc. Ed. Math. Soc., {\bf 2}, 140.}
\\ {[8] Rosen, N. (1930) M.Sc. Thesis, M.I.T., p.20. } \\ {[9] Walker, A.G. (1940) Quar. J. Math., {\bf 11},
81.} \\ {[10] Mikhail, F.I. (1952) Ph.D. Thesis, London
University.} \\ {[11] Hayashi, K. and Bregman, A. (1973) Ann.
Phys. (N.Y.), {\bf 75}, 562.} \\ {[12] McCrea, W.H. and Mikhail,
F.I. (1956) Proc. Roy. Soc. London, {\bf A235}, 11.} \\ {[13]
Bergmann, P.G. and Thomson, R. (1953) Phys. Rev., {\bf 89}, 400.}
\\ {[14] Bilby, B.A., Bullough, R. and Smith, E. (1955) Proc. Roy.
Soc. London, {\bf A231}, 263.} \\ {[15] Utiyama, R. (1956) Phys.
Rev., {\bf 101}, 1597.} \\ {[16] Kibble, T.W.B. (1961) J. Math.
Phys. {\bf 2}, 212.} \\ {[17] Sciama, D.W. (1962) In {\it "Recent
Developments in General Relativity"}

(Festschrift for Infeld), p.415, Pergamon.} \\ {[18] M\o ller, C.
(1961) Math. Fys. Skr. Dan Vid. Selsk., {\bf 1}, 10.} \\ {[19] M\o
ller, C. (1966) Math. Fys. Medd. Dan. Vid. Selsk., {\bf 35}, 3.}
\\ {[20] Mikhail, F.I. (1964) Il Nuovo Cimento, {\bf 32}, 886.} \\
{[21] Hehl, F.W. (1973) Gen. Rel. Grav., {\bf 4}, 333.} \\ {[22]
Wanas, M.I. (1975) Ph.D. Thesis, Cairo University.} \\ {[23]
Mikhail, F.I. and Wanas, M.I. (1977) Proc. Roy. Soc. London, {\bf
A356}, 471.} \\ {[24] Mikhail, F.I. and Wanas, M.I. (1981) Int. J.
Theoret. Phys., {\bf 20}, 671.} \\ {[25] Wanas, M.I. (1981) Il
Nuovo Cimento, B {\bf 66}, 145.}\\ {[26] Wanas, M.I. (1985) Int.
J. Theoret. Phys., {\bf 24}, 639.} \\ {[27]  Wanas, M.I. (1989)
Astrophys. Space Sci.{\bf 154}, 165.} \\ {[28] Hehl, F.W.,
Ne'eman, Y., Nitch, J. and Von der Heyde, P. (1978) Phys. Lett.

{\bf 78B}, 102.} \\ {[29] M\o ller, C. (1978) Math. Fys. Medd.
Dan. Vid. Selsk., {\bf 39}, 1.} \\ {[30] Hayashi, K. and
Shirafuji, T. (1979) Phys. Rev. D{\bf 19}, 3524.} \\ {[31] Nitch,
J. and Hehl, F.W. (1980) Phys. Lett. {\bf 90B}, 98.} \\ {[32]
Muller-Hoissen, F. and Nitch, J. (1983) Phys. Rev. D{\bf 28},
718.}
\\ {[33] Wanas, M.I., Melek, M. and Kahil, M.E. (1995) Astrophys.
Space Sci., {\bf 228}, 273.} \\ {[34] Wanas, M.I. and Kahil, M.E.
(1999) Gen. Rel. Grav., {\bf 31}, 1921.} \\ {[35] Wanas, M.I. and
Melek, M. (1995) Astrophys. Space Sci., {\bf 228}, 277.} \\ {[36]
Wanas, M.I. (2000) Turk. J. Phys., {\bf 24}, 473.\\ {[37] Wanas,
M.I. (1998) Astrophys. Space Sci., {\bf 258}, 237.} \\ {[38] De
Andrade, V.C. and Pereira, J.G. (1998) Gen. Rel. Grav., {\bf 30},
263.} \\ {[39] Wanas, M.I., Melek, M. and Kahil, M.E. (2000) Gravit.
Cosmol., {\bf 6}, 319,

and gr-qc/9812085.} \\  {[40] Wanas, M.I., Melek, M. and Kahil, M.E.
(2000) Proc. Mg9 At3b.  }\\ {[41]} Wanas, M.I. (2000)
Proc-IAU-Symp. \# 201. \\{[42] Hammond, R.T. (1998) Gen. Rel.
Grav. Lett., {\bf 30}, 1803.}
\\ {[43] Hammond, R.T. (1995) Class. Quantum Grav., {\bf 12}, 279.}
\\ {[44] Hehl, F.W., von der Heyde, P., Kerlick, G.D. and Nester,
J.M. (1976)

Rev. Mod. Phys., {\bf 48}, 393.} \\

\end{document}